\documentclass[%
 preprint,
superscriptaddress,
 amsmath,amssymb,
 aps,
prl,
]{revtex4-2}

\usepackage{graphicx}
\usepackage{dcolumn}
\usepackage{bm}
\usepackage[mathlines]{lineno}
\usepackage{appendix}
\usepackage{braket}
\usepackage[dvipsnames]{xcolor}
\usepackage{amsmath,amssymb,amsfonts} 
\usepackage{makecell}
\usepackage{float}
\usepackage{bbm}
\usepackage{nicefrac}
\usepackage{wasysym}
\usepackage[normalem]{ulem}

\usepackage{hyperref}
\hypersetup{
	colorlinks=true,
	linkcolor=blue,
	filecolor=blue,
	urlcolor=blue,
	citecolor=blue,
}

\def\beq{\begin{equation}}
\def\eeq{\end{equation}}
\def\bea{\begin{eqnarray}}
\def\eea{\end{eqnarray}}

\def\kk{{\bm k}}

\begin{document}

\preprint{APS/123-QED}

\title{Two-Terminal Electrical Detection of the N\'eel Vector via Longitudinal Antiferromagnetic Nonreciprocal Transport}

\author{Guozhi Long}
\affiliation{School of Physics, Peking University, Beijing 100871, China}

\author{Hui Zeng}
\affiliation{State Key Laboratory of Low-Dimensional Quantum Physics, Department of Physics, Tsinghua University, Beijing 100084, China}

\author{Mingxiang Pan}
\affiliation{School of Physics, Peking University, Beijing 100871, China}

\author{Wenhui Duan}
\affiliation{State Key Laboratory of Low-Dimensional Quantum Physics, Department of Physics, Tsinghua University, Beijing 100084, China}
\affiliation{Institute for Advanced Study, Tsinghua University, Beijing 100084, China}
\affiliation{Frontier Science Center for Quantum Information, Beijing 100084, China}
\affiliation{Collaborative Innovation Center of Quantum Matter, Beijing 100871, China}

\author{Huaqing Huang}
\email[Corresponding author: ]{huaqing.huang@pku.edu.cn}
\affiliation{School of Physics, Peking University, Beijing 100871, China}
\affiliation{Collaborative Innovation Center of Quantum Matter, Beijing 100871, China}
\affiliation{Center for High Energy Physics, Peking University, Beijing 100871, China}

\date{\today}

\begin{abstract}
We propose a robust two-terminal electrical readout scheme for detecting the N\'eel vector orientation in antiferromagnetic (AFM) materials by leveraging longitudinal nonreciprocal transport driven by quantum metric dipoles. Unlike conventional readout mechanisms, our approach does not require spin-polarized electrodes, tunneling junctions, or multi-terminal geometries, offering a universal and scalable solution for AFM spintronics. As examples, we demonstrate pronounced second-order longitudinal nonlinear conductivity (LNC) in two-dimensional (2D) MnS and 3D CuMnAs, both of which exhibit clear sign reversal of LNC under 180$^\circ$ N\'eel vector reorientation. We show that this LNC is predominantly governed by the intrinsic, relaxation-time-independent quantum metric mechanism rather than the extrinsic nonlinear Drude effect. Our findings provide a practical and material-general pathway for electrically reading AFM memory states, with promising implications for next-generation AFM spintronic technologies.
\end{abstract}

\maketitle

\textit{Introduction.}---Antiferromagnetic (AFM) materials are promising candidates for next-generation spintronic applications due to their ultrafast dynamics, zero stray fields, and robustness against external magnetic perturbations \cite{jungwirth2016antiferromagnetic, RevModPhys.90.015005, han2023coherent,jungwirth2018multiple,nvemec2018antiferromagnetic}. The AFM order parameter, known as the N\'eel vector, serves as a nonvolatile state variable for encoding information in memory devices \cite{RevModPhys.90.015005,jungwirth2018multiple}. A critical prerequisite for realizing such AFM-based technologies lies in achieving efficient electrical detection of the N\'eel vector orientation \cite{RevModPhys.90.015005, han2023coherent, jungwirth2018multiple,nvemec2018antiferromagnetic}. Unlike ferromagnetic (FM) spintronics \cite{PhysRevLett.74.3273, ikeda2010perpendicular, ZHU200636}, where tunneling magnetoresistance (TMR) in two-terminal magnetic tunnel junctions (MTJs) enables the detection of parallel or antiparallel magnetization states [Fig.~\hyperlink{fig-illustration}{1(a)}],
detecting the N\'eel vector in AFMs remains a significant challenge due to the absence of net magnetization \cite{RevModPhys.90.015005}. While recent two-terminal AFM MTJ measurements have leveraged emerging effects such as AFM TMR \cite{shao2021spin, PhysRevX.12.011028, qin2023room, chen2023octupole}, these approaches rely on momentum- or sublattice-dependent spin polarization [Fig.~\hyperlink{fig-illustration}{1(b)}]. Such mechanisms are merely restricted to specific AFM systems, including noncollinear AFMs, collinear altermagnets, and AFMs with strong N\'eel spin current \cite{PhysRevLett.130.216702}.
Recently, several alternative electrical detection schemes—including anisotropic magnetoresistance (AMR) \cite{doi:10.1126/science.aab1031,vzelezny2018spin}, spin Hall magnetoresistance \cite{mahmood2021voltage,PhysRevB.105.094442}, anomalous Hall effect \cite{han2024electrical}, extrinsic \cite{PhysRevLett.115.216806,PhysRevLett.124.067203,du2021nonlinear} and intrinsic nonlinear Hall effects \cite{PhysRevLett.131.056401, PhysRevLett.127.277201, PhysRevLett.127.277202}, and layer Hall effect \cite{gao2021layer,PhysRevLett.133.096803}—have been proposed but have their own limitations. For instance, multi-terminal Hall measurements typically require complex four-electrode configurations [Fig.~\hyperlink{fig-illustration}{1(c)}], complicating experimental implementation and obstructing device miniaturization. Moreover, techniques like AMR and extrinsic nonlinear Hall effects, which are time-reversal-even ($\mathcal{T}$-even), fail to resolve 180$^\circ$ N\'eel vector reversal—a critical requirement for robust binary-state discrimination. These limitations underscore the pressing need for universal, scalable detection methodologies compatible with diverse AFM systems. In this Letter, we demonstrate that longitudinal nonlinear nonreciprocal transport enables efficient N\'eel vector detection using a simplified two-terminal architecture devoid of intricate junction structures or multi-terminal geometry [Fig.~\hyperlink{fig-illustration}{1(d)}].

\begin{figure}
    \hypertarget{fig-illustration}{}
    \centering
    \includegraphics[width=1\columnwidth]{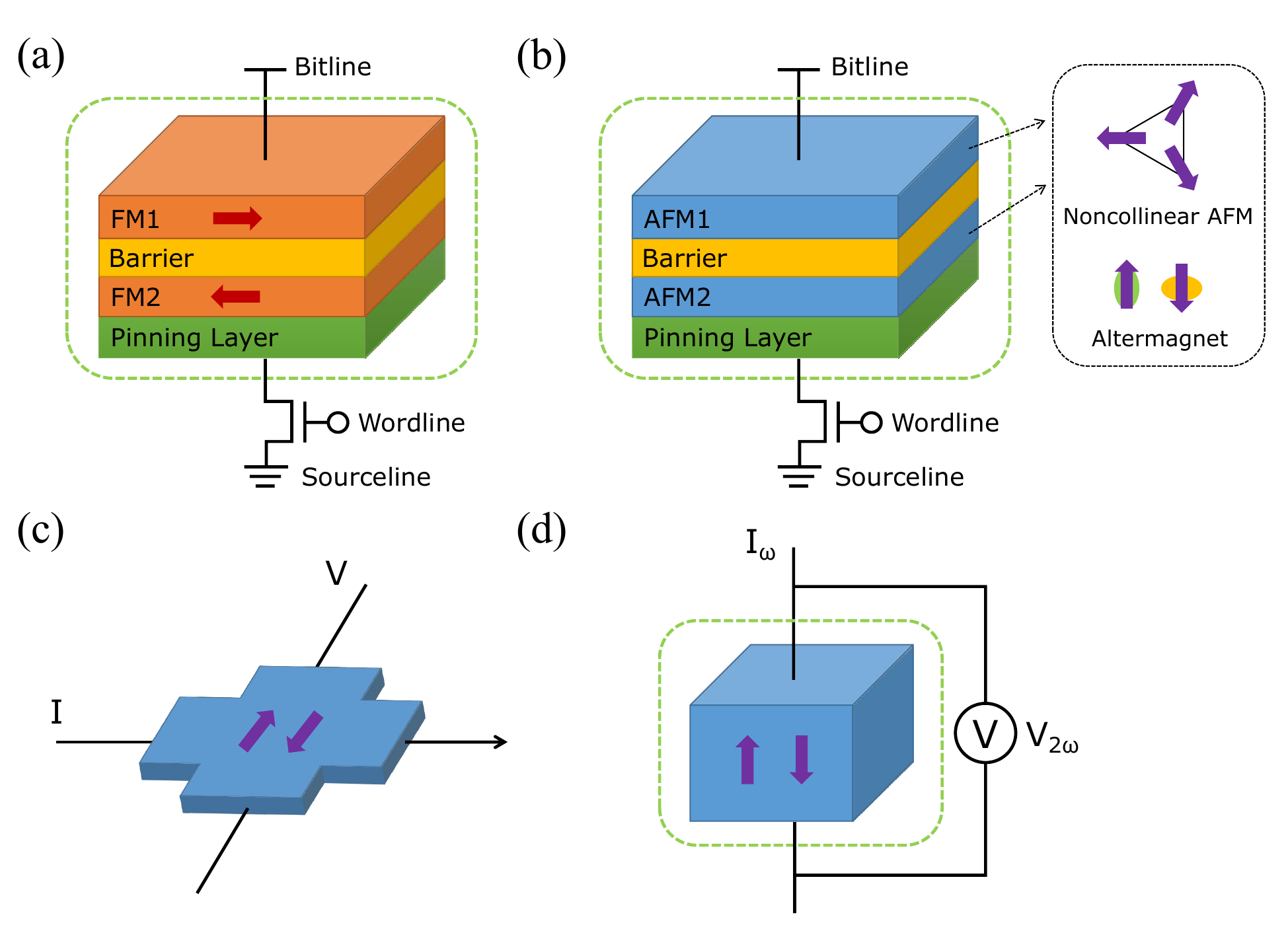}
    \caption{Typical readout schemes for spintronics. (a) Tunneling magnetoresistance (TMR) in magnetic tunnel junction (MTJ) with two terminals for detecting the magnetic moment (i.e., state variable to encode the information). (b) Antiferromagnetic MTJ for detecting the N\'eel vector (i.e., state variable in antiferromagnets). (c) Hall measurement with four terminals for detecting the N\'{e}el vector. (d)  Detection of the AFM N\'eel vector via longitudinal nonreciprocal transport without intricate junction structures or multi-terminal geometry.
    }
\end{figure}

Longitudinal nonlinear transport has been observed in non-centrosymmetric polar and chiral conductors \cite{doi:10.1126/sciadv.1602390, nadeem2023superconducting, tokura2018nonreciprocal, ideue2017bulk, li2021nonreciprocal, PhysRevLett.132.046303, PhysRevResearch.3.033253,PhysRevB.106.115202,PhysRevLett.129.276601,PhysRevB.104.054429,PhysRevLett.128.176602,zhang2022controlled}, leading to nonreciprocal charge transport with diode-like I-V characteristics, similar to those in semiconductor p-n junctions \cite{tokura2018nonreciprocal,annurev-conmatphys-032822-033734,annurev-conmatphys-060220-100347}. Under an external magnetic field $B$, these materials can exhibit magnetochiral anisotropy and unidirectional magnetoresistance, where the resistance changes sign upon reversing either the magnetic field or electric current, effectively acting as a magnetic diode \cite{ideue2017bulk, annurev-conmatphys-042924-123620}. Recently, a similar nonreciprocal effect driven by intrinsic magnetic order has been reported in $\mathcal{T}$-breaking FM and AFM materials \cite{godinho2018electrically,zhang2022controlled, gao2024antiferromagnetic, AdvMat_CrSBr}. The underlying mechanisms for these nonreciprocal responses include the external nonlinear Drude effect \cite{PhysRevLett.129.276601, PhysRevLett.133.096802, PhysRevLett.134.046801, PhysRevResearch.2.033100, PhysRevResearch.2.043081, PhysRevX.11.011001, PhysRevB.104.054429, PhysRevLett.117.127202} and skew scattering \cite{doi:10.1126/sciadv.aay2497, he2022graphene}, both dependent on the relaxation time $\tau$, as well as the intrinsic quantum metric effect, which is independent of $\tau$ \cite{PhysRevLett.132.026301, PhysRevB.108.L201405}. Moreover, the interplay of these mechanisms has inspired an increasing interest in nonlinear nonreciprocal transport \cite{gong2025Haizhou_Lu, huang2024scalinglaw, JPSJ.91.014701, PhysRevB.110.245406}.

In this Letter, we recognize that an AFM nonreciprocal transport contributes
to the $\mathcal{T}$-odd second-order longitudinal nonlinear conductivity (LNC) and can be used to detect the reversal of the N\'eel vector. Taking the 2D MnS and 3D CuMnAs as prototype examples, we show that the LNC is pronounced in these materials and exhibits sensitive dependence on the N\'eel vector orientation with a $2\pi$ periodicity, which suggests an efficient way to detect the N\'eel vector. Moreover, the quantum-metric-induced LNC, which mainly stems from the band near degeneracy, is predominant over the second-order Drude contribution, which solely relies on the band asymmetry between opposite momenta and the relaxation time. Our research highlights the role of quantum metrics in AFM nonreciprocal effect and provides a simple yet effective means for readout of the AFM state without the reliance on multi-terminal geometries, intricate junction structures, or spin-polarized electrodes.

\textit{Longitudinal nonlinear conductivity.}---In the study of nonlinear transport, a longitudinal current is predicted to be generated by a harmonically oscillating electric field $E_\alpha=\mathrm{Re}\{\mathcal{E}_\alpha e^{i\omega t}\}$ in noncentrosymmetric systems. The response current up to second order reads $j^\alpha = \mathrm{Re} \{ j^\alpha_{0} + j^\alpha_{2\omega} e^{2i\omega t} \} $, where a rectified current $j^\alpha_{0} = \sigma^{\alpha\alpha\alpha}_0(\omega)\, \mathcal{E}_\alpha \mathcal{E}_{\alpha}^{*}$ and a second-harmonic current $j^\alpha_{2\omega} =\sigma^{\alpha\alpha\alpha}_{2\omega} (\omega)\, \mathcal{E}_\alpha \mathcal{E}_\alpha$ depend on the nonlinear conductivity tensors. In the $\omega\rightarrow0$ limit, the longitudinal conductivity $\sigma^{\alpha\alpha\alpha}_{0,2\omega}(\omega)\rightarrow (\sigma^{\alpha\alpha\alpha}_\mathrm{Mag} + \sigma^{\alpha\alpha\alpha}_\mathrm{Drude})/2$ has two contributions from the band-normalized quantum metric dipole induced nonreciprocal magnetoresistance ($\sim \tau^0$) \cite{PhysRevLett.132.026301, PhysRevB.108.L201405} and the nonlinear Drude conductivity ($\sim\tau^2$) \cite{PhysRevResearch.2.033100, PhysRevResearch.2.043081, PhysRevX.11.011001, PhysRevB.104.054429}, respectively:
\begin{eqnarray}
    \sigma^{\alpha \beta \gamma}_{\mathrm{Mag}} & = &\frac{e^3}{\hbar} \sum_n \int_k \frac{\partial f_n}{\partial \epsilon_n} v_n^\alpha G_n^{\beta \gamma},\label{mag_formula}\\
    \sigma^{\alpha \beta \gamma}_{\mathrm{Drude}}&=&\frac{e^3 \tau^2}{\hbar^3} \sum_n \int_k \frac{\partial f_n}{\partial \epsilon_n} v_n^\alpha \frac{\partial^2 \epsilon_n}{\partial k_\beta \partial k_\gamma}.
\label{drude_formula}
\end{eqnarray}
Here, 
the integral $\int_k \equiv \int_\mathrm{BZ}\frac{d^D k}{(2\pi)^D}$ denotes integration over the Brillouin zone (BZ) in \( D \) dimensions, $v_n^\alpha(\kk) = \frac{\partial \epsilon_n(\kk)}{\partial k_\alpha}$ is the velocity of the \( n \)-th band with energy $\epsilon_n(\bm{k})$, and $f_n$ 
is the Fermi-Dirac occupation function for the \( n \)-th band. 
\( G_n^{\alpha \beta} \) is the energy-normalized quantum metric, which is also described as the Berry connection polarizability:
\begin{equation}
    G_n^{\alpha \beta}(\kk) = 2 \mathrm{Re} \sum_{m \neq n} \frac{A_{nm}^\alpha(\kk) A_{mn}^\beta(\kk)}{\epsilon_n(\kk) - \epsilon_m(\kk)},
\end{equation}
where \( A_{mn}^\alpha(\kk) = i \langle u_{m\kk} | \partial_{ k_\alpha} | u_{n\kk} \rangle \) is the interband Berry connection with \( |u_{n\kk}\rangle \) the periodic part of the Bloch wave function. Note that in addition to the LNC, $\sigma^{\alpha \beta \gamma}_{\mathrm{Mag}}$ also contributes to transverse responses, but it is distinct from the intrinsic nonlinear Hall conductivity $\sigma^{\alpha\beta\gamma}_\mathrm{Hall} =-\frac{e^3}{\hbar} \sum_n \int_k f_n \left[\partial_{k_\alpha} G_n^{\beta \gamma} -
\frac{1}{2}  \left(\partial_{k_\beta} G_n^{\gamma \alpha }
+\partial_{k_\gamma} G_n^{\beta \alpha}\right)\right]$,
which vanishes whenever $\alpha=\beta=\gamma$ and has been studied in previous works \cite{PhysRevLett.127.277201, PhysRevLett.127.277202, PhysRevLett.131.056401}.

\begin{figure*}
    \hypertarget{fig-MnS-Mag}{}
    \centering
    \includegraphics[width=1.0\textwidth]{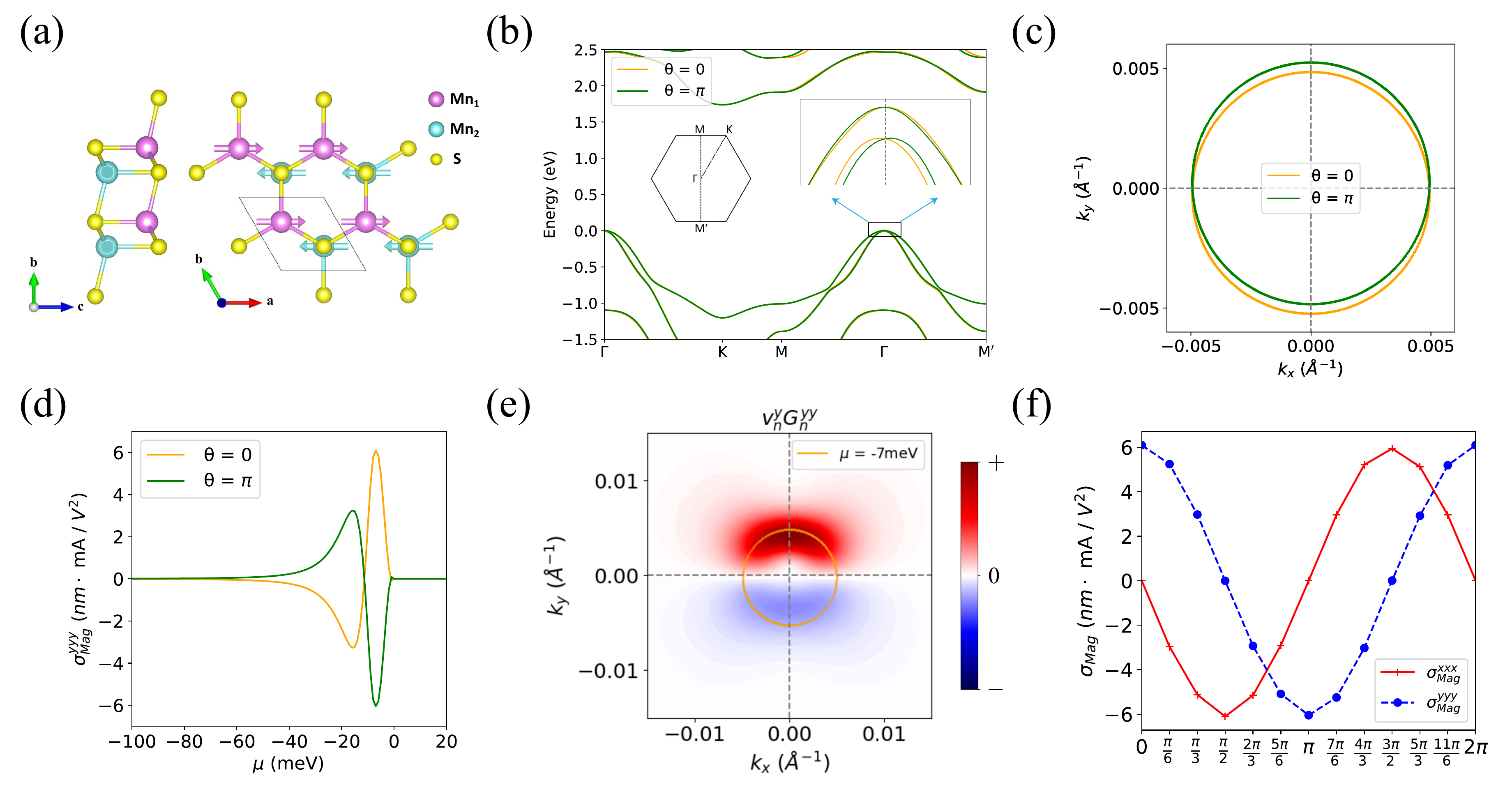}
    \caption{(a) The atomic structure of MnS $(\theta = 0)$ is depicted from both the side and top views, with Mn atoms possessing opposite magnetic moments represented in two distinct colors. (b)
    Band structure of MnS with $\theta = 0$ (orange) and $\theta=\pi$ (green). The Fermi level is set at the valence band maximum. The insert shows the BZ and the enlarged plot of the top two valence bands around the $\Gamma$ point. (c) Iso-energy surfaces at $\mu=-7$ meV for MnS with $\theta = 0$ and $\pi$, which exhibit opposite asymmetric distributions along the $k_y$ direction. (d) The quantum metric-induced LNC $\sigma_\mathrm{Mag}^{yyy}$ of MnS with $\theta = 0$ and $\pi$. (e) Distribution of $\lambda_n^{yyy}(\kk)=v^y_nG_n^{yy}$ around $\Gamma$ for the top valence band of MnS with $\theta=0$. The orange line plots the iso-energy surfaces at $\mu=-7$ meV. (f) $\sigma_\mathrm{Mag}^{yyy}$ and $\sigma_\mathrm{Mag}^{xxx}$ of MnS at $\mu=-7$ meV when the N\'eel vector (denoted by $\theta$) rotates in the $x$-$y$ plane. }
\end{figure*}

\textit{Symmetry analysis.}---To unravel the nonreciprocal nature of LNC in AFM materials, we analyze symmetry conditions that permit a finite LNC. Since the LNC arises from an integral over the symmetric BZ, as described in Eqs.~\eqref{mag_formula} and \eqref{drude_formula}, a nonvanishing LNC essentially requires an asymmetric distribution of the physical quantities associated with band dispersion or Bloch wavefunctions. In particular, the quantum metric-induced LNC, $\sigma^{\alpha\beta\gamma}_\mathrm{Mag}$, 
vanishes in the presence of inversion $\mathcal{P}$ or time-reversal $\mathcal{T}$ symmetry. Because these symmetries relate states at $\kk$ and $-\kk$, enforcing antisymmetry in the quantum-metric dipole $\lambda^{\alpha\beta\gamma}_n(\kk) =v^{\alpha}_n G_n^{\beta\gamma}$, such that $\lambda^{\alpha\beta\gamma}_n(\kk) \rightarrow -\lambda^{\alpha\beta\gamma}_n(-\kk)$. As a result, a finite $\sigma^{\alpha\beta\gamma}_\mathrm{Mag}$ requires the breaking of both $\mathcal{P}$ and $\mathcal{T}$, though it is allowed in antiferromagnets preserving the combined $\mathcal{PT}$ symmetry. Moreover, the $\mathcal{T}$-odd nature of $\sigma^{\alpha\beta\gamma}_\mathrm{Mag}$ allows it to distinguish between N\'eel vector reversals that are related by $\mathcal{T}$. Additional symmetry constraints from crystal symmetries can further suppress specific components of LNC. For example, mirror symmetry $\mathcal{M}_\alpha$, which maps $k_\alpha\rightarrow-k_\alpha$, forbids LNC along the $\alpha$ direction, i.e., $\sigma_\mathrm{Mag}^{\alpha\alpha\alpha}=0$, although other components may still be symmetry-allowed. A complete symmetry analysis of the LNC is provided in Supplemental Material (SM) \footnote{\label{fn}See Supplemental Material at http://link.aps.org/supplemental/xxx, for details on the quantum metric-induced nonlinear conductivity formalism, symmetry analysis (including a complete list of spin point groups permitting LNC), effective model analysis of MnS, first-principles computational methods, extended results for 2D MnS and 3D CuMnAs, and supplementary data for additional antiferromagnetic systems such as 2D CrSBr bilayers and 3D Mn$_2$Au, which include Refs.~\cite{PhysRevLett.132.026301, PhysRevResearch.2.033100, PhysRevResearch.2.043081, PhysRevX.11.011001, PhysRevB.104.054429, PhysRevLett.133.096802, PhysRevX.12.021016, PhysRevLett.131.056401, PhysRevLett.127.277201, PhysRevLett.127.277202, KRESSE199615,PhysRevLett.77.3865, MOSTOFI2008685}.}. Furthermore, since the prerequisite of asymmetry for a nonzero LNC 
can also appear in magnetic materials with negligible relativistic SOC, we complement our analysis with constraints derived from the spin point group \cite{PhysRevX.12.021016, PhysRevX.14.031039, PhysRevX.14.031037}, extending previous studies based on the magnetic point group \cite{PhysRevLett.133.096802}.

\textit{Longitudinal nonreciprocal transport in 2D AFM MnS.}---The 2D MnS depicted in Fig.~\hyperlink{fig-MnS-Mag}{2(a)} forms a buckled bilayer honeycomb lattice structure, where two manganese (Mn) atoms occupy opposite sublattices on the top and bottom layers (Mn$_1$ and Mn$_2$), constituting the AFM ordering \cite{acsnano.1c05532, PhysRevB.106.085410}. The N\'{e}el vector $\boldsymbol{N}$ is defined as the difference of the magnetic moments between Mn$_1$ and Mn$_2$ atoms within the unit cell, and its orientation is indicated by the polar angle $\theta$ relative to the $x$-axis [see Fig.~\hyperlink{fig-MnS-Mag}{2(a)}]. Our first-principles calculations \footnotemark[\value{footnote}] demonstrate that the band structure of MnS depends on the N\'eel vector orientation, which exhibits an opposite band asymmetry along the $k_y$ direction for a 180$^\circ$ reversal of the N\'eel vector ($\theta=0$ and $\pi$), as shown in Fig.~\hyperlink{fig-MnS-Mag}{2(b,c)}. Notably, there is a minor energy splitting between the top two valence bands at $\Gamma$ due to the weak SOC, forming a nearly degenerate point (NDP), as shown in the insert of Fig.~\hyperlink{fig-MnS-Mag}{2(b)}. As will be discussed later, this NDP makes significant contributions to the nonlinear transport.

Subsequently, we calculated the quantum metric-induced LNC of MnS. For $\theta=0$ or $\pi$, the system belongs to the magnetic point group $2'/m$, which includes the mirror reflection across the $x$-axis $\mathcal{M}_x$, and a two-fold rotation about the $x$-axis followed by time reversal $C_{2x}\mathcal{T}$. Due to the symmetry constraint, only the $\sigma^{yyy}_{\mathrm{Mag}}$ component is allowed. As illustrated in Fig.~\hyperlink{fig-MnS-Mag}{2(d)}, $\sigma^{yyy}_{\mathrm{Mag}}$ for $\theta=0$ and $\pi$ are exactly opposite in sign, demonstrating the AFM nonreciprocal effect which can distinguish the reversal of the N\'eel vector. Remarkably, the LNC develops two peaks with opposite signs for hole doping, and the peak magnitudes at $\mu=-7$ meV and $\mu=-18$ meV are on the order of $\mathrm{nm}\cdot \mathrm{mA}/\mathrm{V}^2$. In addition, we noted that the transverse response $\sigma_\mathrm{Mag}^{yxx}$ and the INHE $\sigma_\mathrm{Hall}^{yxx}$ \cite{PhysRevLett.131.056401} resemble the behavior of LNC with similar magnitudes (see Fig. S2 in Supplementary Materials (SM)\footnotemark[\value{footnote}]). This is because all these effects arise from the augmented quantum metric dipole $\lambda_n^{\alpha\beta\gamma}(\kk)$ around the NDPs at $\Gamma$ [see Fig.~\hyperlink{fig-MnS-Mag}{2(e)} for the example of $\lambda_n^{yyy}$].

Next, we discuss the strong dependence of LNC on the orientation of the N\'eel vector. As shown in Fig.~\hyperlink{fig-MnS-Mag}{2(f)}, upon rotating the N\'{e}el vector in the $x$-$y$ plane, the variation of $\sigma_{\mathrm{Mag}}^{yyy}$ and $\sigma_{\mathrm{Mag}}^{xxx}$ exhibit a $2\pi$ periodicity with trigonometric dependences on $\theta$. Therefore, the measurement of LNC is capable of fully mapping out the N\'eel vector orientation, indicating LNC as a powerful electric detection scheme for AFM spintronics. This dependence comes about because the band structure and quantum metric dipole depend on the orientation of the N\'eel vector. Based on a simple $k \cdot p$ model around $\Gamma$ for the top two valence bands, we treat the N\'eel vector rotation as an effective inverse coordination transformation (see SM~\footnotemark[\value{footnote}]). The angle ($\theta$) dependence in the new coordinates $(x',y')$ is obtained as $\sigma^{x^\prime x^\prime x^\prime}_\mathrm{Mag} \approx -\sin \theta \cdot \sigma^{yyy}_\mathrm{Mag}$ and $\sigma^{y^\prime y^\prime y^\prime}_\mathrm{Mag} \approx \cos \theta \cdot \sigma^{yyy}_\mathrm{Mag}$, verifying the behaviors of LNC in Fig.~\hyperlink{fig-MnS-Mag}{2(f)}.

\begin{figure}
    \hypertarget{fig-MnS-Drude}{}
    \centering
    \includegraphics[width=0.65\columnwidth]{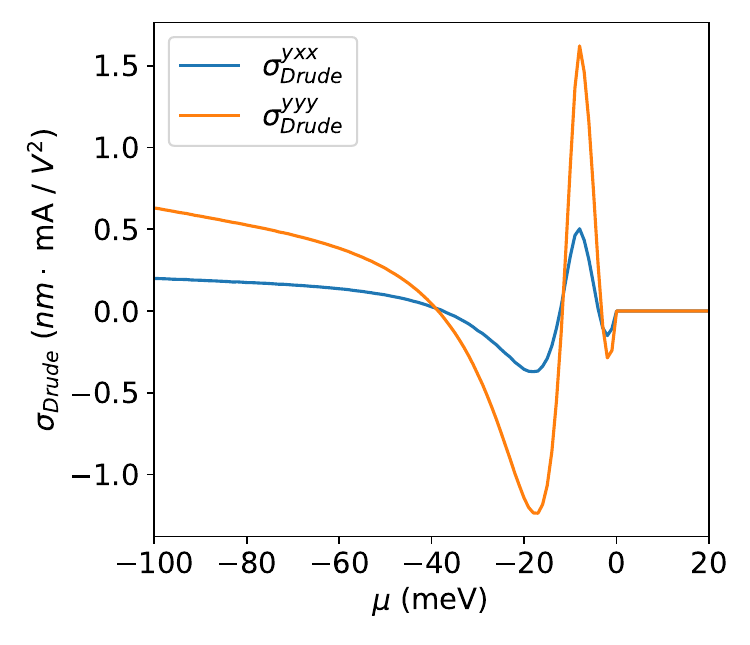}
    \caption{The transverse and longitudinal nonlinear Drude conductivities $\sigma_\mathrm{Drude}^{yxx}$ and $\sigma_\mathrm{Drude}^{yyy}$ for MnS with $\theta=0$.}
\end{figure}

\textit{Nonlinear Drude conductivity.}---We further evaluate the contribution of the nonlinear Drude term to both the transverse and longitudinal nonlinear conductivity. Unlike the quantum metric-induced term $\sigma_\mathrm{Mag}^{\alpha\beta\gamma}$, the nonlinear Drude conductivity $\sigma_\mathrm{Drude}^{\alpha\beta\gamma}$ depends solely on the asymmetric band dispersion and scales with $\tau^2$.  In moderately conducting materials, $\sigma_\mathrm{Mag}$ is generally expected to dominate over $\sigma_\mathrm{Drude}$. To quantify this, we estimate the nonlinear Drude conductivity in MnS using a typical relaxation time of $\tau = 60$ femtoseconds \cite{zhou2020high}.  As shown in Fig.~\hyperlink{fig-MnS-Drude}{3}, the transverse component $\sigma_{\mathrm{Drude}}^{yxx}$ is smaller than the longitudinal component $\sigma_{\mathrm{Drude}}^{yyy}$, and the latter remains 
significantly smaller than the quantum metric-induced contribution $\sigma_\mathrm{Mag}^{yyy}$ in MnS.

\textit{Longitudinal nonreciprocal transport in 3D AFM Tetragonal CuMnAs.}---Beyond 2D AFM, we further explore longitudinal nonreciprocal transport in 3D AFM materials. A notable example is tetragonal CuMnAs, a metastable phase successfully synthesized via molecular beam epitaxy on GaAs and GaP substrates \cite{Wadley2013tetragonal}. This compound exhibits a high N\'eel temperature of $ 480\pm5$ K \cite{saidl2017optical}, making it a promising candidate for room-temperature AFM spintronic applications. Fig.~\hyperlink{fig-CuMnAs}{4(a)} illustrates its atomic structure: each unit cell comprises two Mn planes with magnetic moments lying in-plane and ordered antiferromagnetically between the planes. The N\'eel vector $\bm{N}$, defined as the difference between the magnetic moments of Mn$_1$ and Mn$_2$ atoms, can also be parameterized by the polar angle $\theta$.

In Fig.~\hyperlink{fig-CuMnAs}{4(b)}, we present the dependence of \( \sigma_\mathrm{Mag}^{yyy} \) on the chemical potential. With electron doping, \( \sigma_\mathrm{Mag}^{yyy} \) is significantly enhanced, reaching approximately \(-1.0\) mA/V\(^2\) when the chemical potential shifts to \( \mu = 22 \) meV.
The significant LNC can be anticipated from the quantum metric dipole distribution plotted on the isoenergy surface of $\mu=22$ meV in the $k_x = 0$ plane of the BZ [bottom panel of Fig.~\hyperlink{fig-CuMnAs}{4(c)}]. Because of the NDPs with obvious asymmetries in the $\mathrm{Y^{\prime}-\Gamma-Y}$ line at that energy [the top panel of Fig.~\hyperlink{fig-CuMnAs}{4(c)}], the isoenergy surface has a significant asymmetric quantum metric dipole around these NDPs. 
Figure~\hyperlink{fig-CuMnAs}{4(d)} shows the variation of \( \sigma_\mathrm{Mag}^{yyy} \) and \( \sigma_\mathrm{Mag}^{xxx} \) as the N\'eel vector \( \bm{N} \) rotates within the $x$-$y$ plane. Notably, 
both components exhibit the \( 2\pi \) periodicity and obey the $\mathcal{T}$-odd constraint, \( \sigma_\mathrm{Mag}^{\alpha\alpha\alpha}(\theta) = -\sigma_\mathrm{Mag}^{\alpha\alpha\alpha}(\theta + \pi) \). Consequently, the reorientation of \( \bm{N} \) can be effectively monitored through LNC measurements.

\begin{figure}
    \hypertarget{fig-CuMnAs}{}
    \centering
    \includegraphics[width=1.0\columnwidth]{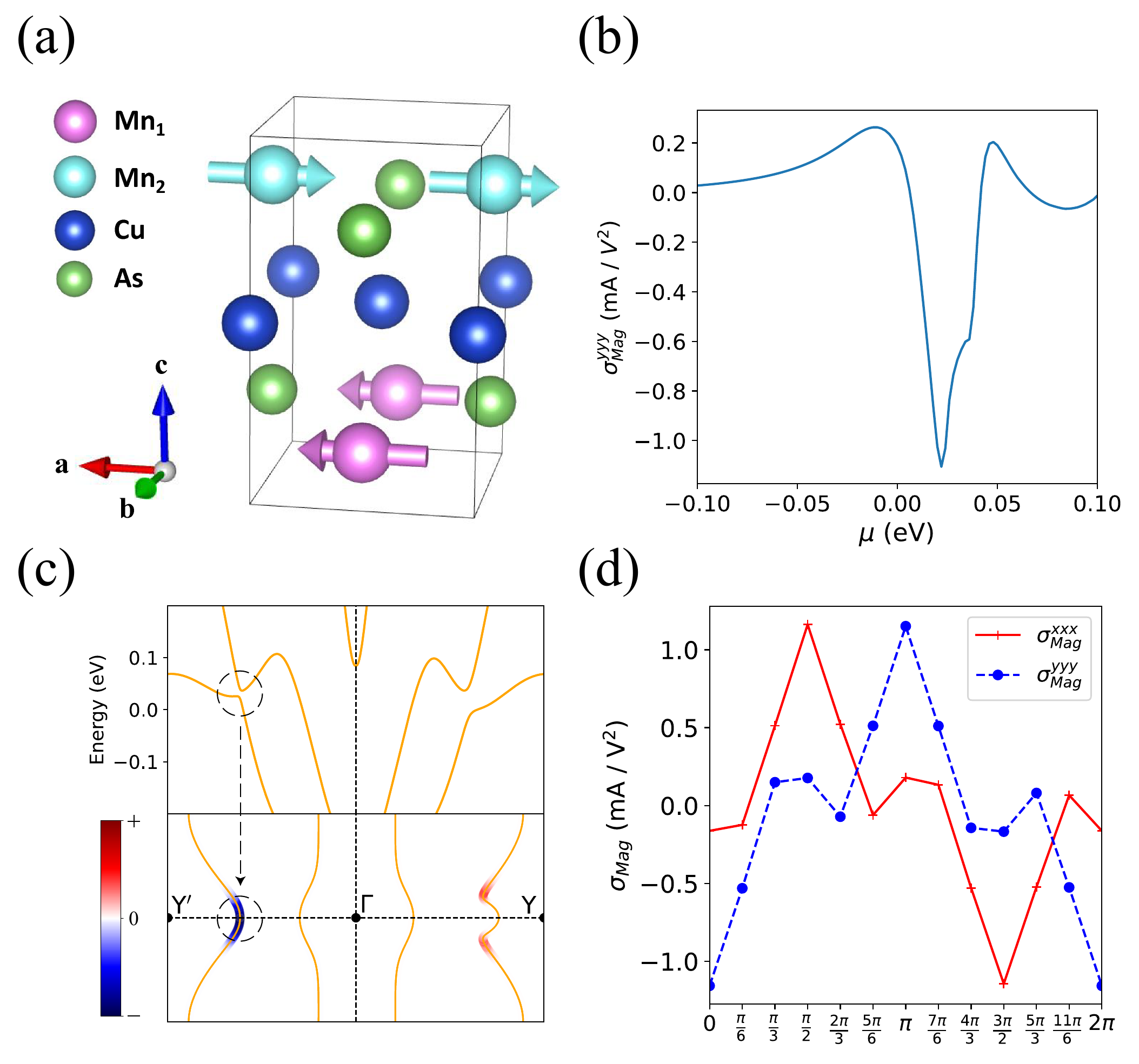}
    \caption{(a) Atomic structure of tetragonal CuMnAs. (b) Quantum-metric induced LNC $\sigma_\mathrm{Mag}^{yyy}$ of CuMnAs with $\theta=0$. (c) The upper panel shows the band structure along the $\mathrm{Y^{\prime}-\Gamma-Y}$ line. The lower panel shows the distribution of $\sum_n[\frac{\partial f_n}{\partial \epsilon_n}\lambda_n^{yyy}(\kk)]$ (color map) on the isoenergy surface of $\mu=22$ meV (orange lines) in the $k_x=0$ plane of the BZ.
    (d) $\sigma_\mathrm{Mag}^{yyy}$ and $\sigma_\mathrm{Mag}^{xxx}$ of CuMnAs at $\mu=22$ meV when the N\'eel vector (denoted by $\theta$) rotates in the $x$-$y$ plane.}
\end{figure}

\textit{Discussion and summary.}---Recent experimental studies on tetragonal CuMnAs have demonstrated the detection of N\'eel vector reversal by observing the sign change in second-order conductivity \cite{godinho2018electrically}. Furthermore, both the angular dependence and the measured magnitude of LNC align qualitatively with our theoretical predictions~\footnotemark[\value{footnote}]. These findings underscore LNC as a viable and practical method for probing \( \bm{N} \), addressing a critical challenge in the readout schemes of antiferromagnetic spintronic devices.

In summary, we have proposed a highly efficient two-terminal electrical method for detecting the orientation of the N\'{e}el vector by harnessing quantum metric-induced nonreciprocal transport in AFMs. Using 2D MnS and 3D CuMnAs as representative examples, we demonstrate that the LNC changes sign upon a 180$^\circ$ reversal of the N\'{e}el vector, corresponding to two distinct memory states in AFM spintronic devices. Given the broad experimental accessibility of abundant AFM materials, the proposed approach is applicable to a wide range of compensated AFMs, including MnBi$_2$Te$_4$ \cite{gao2024antiferromagnetic, PhysRevLett.132.026301, PhysRevLett.129.276601}, Mn$_2$Au \cite{PhysRevApplied.9.064040,PhysRevB.99.140409,PhysRevLett.127.277202}, $\varepsilon$-Fe$_2$O$_3$ \cite{adma.202209465,PhysRevLett.133.096802}, Cr$_2$O$_3$ \cite{mahmood2021voltage,PhysRevLett.133.096803}, and van der Waals AFMs such as CrSBr \cite{AdvMat_CrSBr,das2025surface}, CrI$_3$ \cite{Huang2018CrI3,Sun2019Giant}, and MnPS$_3$ \cite{Ni2021imaging,PhysRevLett.124.027601}, some of which are further demonstrated in the SM~\footnotemark[\value{footnote}]. Beyond N\'{e}el vector detection, the intrinsic nonlinearity associated with AFM nonreciprocal transport also presents promising opportunities for wireless radio-frequency rectification and terahertz photodetection \cite{ideue2017bulk, kumar2021room, Kumar2024quantum, Zeng2024Te.rectification, doi:10.1126/sciadv.aay2497, he2022graphene, pnas.2100736118, PhysRevB.110.075122, PhysRevB.110.155122, guo2022ultrasensitive,adma.202400729}, paving the way for next-generation spintronic technologies.

\begin{acknowledgments}
This work is supported by the National Key R\&D Program of China (Grant No. 2021YFA1401600), the National Natural Science Foundation of China (Grant No. 12474056), and the 2022 basic discipline top-notch students training program 2.0 research project (Grant No. 20222005).
H.Z. and W.D. acknowledge support from the Basic Science Center Project of NSFC (Grant No. 52388201), the Ministry of Science and Technology of China, and the Innovation Program for Quantum Science and Technology (Grants No. 2023ZD0300500).
The work was carried out at the National Supercomputer Center in Tianjin, and the calculations were performed on Tianhe new generation supercomputer. The high-performance computing platform of Peking University supported the computational resources.
\end{acknowledgments}


%

\end{document}